\documentstyle[12pt]{article}

\begin{document}

\setcounter{page}{1}

\pagestyle{plain}

\begin{center}
\Large{\bf Braneworld Mimetic Cosmology }\\
\small \vspace{1cm} {\bf Naser Sadeghnezhad\footnote{nsadegh@stu.umz.ac.ir}}\quad and \quad  {\bf Kourosh
Nozari\footnote{knozari@umz.ac.ir(Corresponding Author)}}\\
\vspace{0.5cm} Department of Physics, Faculty of Basic Sciences,
University of Mazandaran,\\
P. O. Box 47416-95447, Babolsar, IRAN
\end{center}

\begin{abstract}
We extend the idea of mimetic gravity to a Randall-Sundrum II braneworld model. As for the 4-dimensional mimetic gravity, we isolate the conformal degree of freedom of 5-dimensional gravity in a covariant manner. We assume the bulk metric to be made up of a non-dynamical scalar field $\Phi$ and an auxiliary metric $\tilde{{\cal{G}}}_{AB}$ so that
${\cal{G}}_{AB}= \tilde{{\cal{G}}}^{CD}\,\Phi_{,C}\,\Phi_{,D}\,\tilde{{\cal{G}}}_{AB}$ where $A, B, ...$ are the bulk spacetime indices. Then we show that the induced conformal degree of freedom on the brane as an induced scalar field, plays the role of a mimetic field on the brane. In fact, we suppose that the scalar degree of freedom which mimics the dark sectors on the brane has its origin on the bulk scalar field, $\Phi$. By adopting some suitable mimetic potentials on the brane, we show that this brane mimetic field explains the late time cosmic expansion in the favor of observational data: the equation of state parameter of this field crosses the cosmological constant line in near past from quintessence to phantom phase in a redshift well in the range of observation. We show also that this induced mimetic scalar field has the capability to explain initial time cosmological inflation. We study parameter space of the models numerically in order to constraint the models with Planck2015 data set. 
\end{abstract}
\newpage

\section{Introduction}

Recent observational data are in the favor of an accelerating phase
of the Universe expansion. To explain this late time cosmic speed up, one can
follow two main approaches. The first approach is to add some
mysterious components (dubbed dark energy) in the energy-momentum
sector of the Einstein field equations. The cosmological constant is a possible candidate
for dark energy. However, some yet unsolved problems
such as unknown origin, lake of dynamics and a huge amount of fine
tuning for its magnitude, have led cosmologist to propose other
candidates for a dynamical dark energy component. In this respect, several types of scalar fields such as
quintessence, phantom, tachyon and k-essence are considered as possible
candidates for dark energy so far; see for instance ~\cite{Ratra,Wetterich,Caldwell1,Caldwell2,Nojiri88,Copeland,Padmanabhan1,Sen,Sen1,Nozari2,
Armendariz-Picon,Chiba}. The second approach is modification of
the geometric (gravitational) sector of the Einstein field equations. In comparison with ``dark
energy'', this modification is called usually as the ``dark geometry''. Modification of the geometric sector is accomplished in several ways like as
modification of the Einstein-Hilbert action by replacing the Ricci scalar with a generic function of this scalar as
$f(R)$~\cite{Sotiriou,Nojiri,Capozziello}, considering the
Gauss-Bonnet term or even higher order derivative terms in the spirit of general scalar-tensor theories in the action of the
model~\cite{Zwi85,Bou85,Noj05,Noj07,Guo09,And07,Noz09b,Noz09c} and
adopting braneworld scenarios~\cite{Lid04,Dva00,Dva01a,Lue06,Laz04,RS1,RS2}. Our
attention in this paper is paid to an extension of the braneworld scenarios in the spirit of recently proposed mimetic gravity.

From superstring theories, to have a worthwhile theory
our observed universe should be a membrane embedded in a higher dimensional spacetime
called the bulk. Extra spatial dimensions should be compactified on the scales that compared to
usual four spatial dimensions are so small~\cite{Pol98,H96a,H96b}. In this framework, gravity propagates
through the entire spacetime, whereas the ordinary matters are
trapped on the brane. DGP (Dvali-Gabadadze-Porrati)~\cite{Dva00,Dva01a} and Randall-Sundrum (RS) (I and
II)~\cite{RS1,RS2} braneworld models are the ones in which the
universe is considered to be a 5D spacetime and our 4D world is embedded in a 5D bulk. The DGP setup is based on the
modification of the gravitational sector of the theory over large
distances in an induced gravity perspective. In this model, the bulk
is a flat Minkowski spacetime. On the other hand, in RS (I and II)
models the bulk is $AdS_{5}$. RS I model, which was proposed to solve
the hierarchy problem, consists of two Minkowski brane embedded in
$AdS_{5}$ bulk. In this model, the standard matters are confined on
the brane with negative tension (embedded at $y=y_{c}$) and then
gravity is confined on the hidden brane with positive tension
(embedded at $y=0$). Gravity leaks off the brane and through the bulk
reaches to our brane. However, in this model there are some
problems like as the stabilization of the Radion and the lack
of the acceptable cosmology. In the second RS model (RS II model)
$y_{c}$ is considered to be infinite and our observed universe is
located on the brane with positive tension, embedded at
$y=0$~\cite{RS2,Gum}. In this model, the spacetime is effectively
compactified within the curvature radius $l$ of the $AdS_{5}$ bulk
and in length scale larger than $l$, the 4-dimensional Einstein
gravity is effectively recovered on the brane. Actually, in this
model since there are a negative cosmological constant in the bulk
and a positive tension on the brane (in addition to the ordinary
matter), it is possible to cancel out the bulk energy's contribution on
the brane and get the standard 4-dimensional Friedmann equation at
late time, corresponding to the low energy scales~\cite{RS2,Gum}.
RS II model is one of the interesting barneworld models (specially for a viable initial time cosmology) and some
authors have studied its cosmological
aspects in details (see for instance ~\cite{Kal99,Bin00,Bri02,Lan01,Koy02,Fla01,Gar00,Maa00,Lan00,Koy00}).
In this regard, the authors of paper~\cite{Bin00} have studied an
evolving universe with any types of matter on the brane and a
cosmological constant in the bulk, and solved the 5D Einstein's
field equations. In Refs.~\cite{Dav01,Noz12b,Noz13d} the authors
have considered a scalar field in the bulk and studied the
cosmological solutions both in bulk and brane in their setup. The authors of
Ref.~\cite{Him01} have studied the cosmological inflation driven by a
dilaton-like gravitational field in the bulk. By modeling the
effective potential of the gravitational scalar field, they have
obtained the solution of the bulk field's equation and found that
the solution gives slow-roll inflation on the brane. In
Ref.~\cite{Him02} the dynamics of the bulk scalar field in more
general situations has been discussed and it has been shown that
there is a simple relation between the 5D potential and the
effective 4D potential on the brane. This feature is an essential preliminary in our work.

In 2013, an interesting model for gravity (the so called \emph{mimetic gravity}) has been proposed which
can explain the origin of the dark sectors in a fascinating manner ~\cite{Chamseddine}. In
the original model, a free and non-dynamical scalar field $\phi$
and an auxiliary metric $\tilde{g}_{\mu\nu}$ are ingredients of the
physical metric $g_{\mu\nu}$ with the following definition
\begin{eqnarray}\label{eq1}
g_{\mu\nu}= \tilde{g}^{\alpha\beta}\,\phi_{,\alpha}\,\phi_{,\beta}
\,\tilde{g}_{\mu\nu}\,.
\end{eqnarray}
This model respects the conformal symmetry as an internal degree of
freedom and the scalar field encodes the conformal mode of the
gravity. If we perform a Weyl transformation of the auxiliary
metric, the physical metric is invariant. When we define the
physical metric as in (\ref{eq1}), an extra longitudinal mode of
the gravitational field would be appeared in the equations of motion
which ensures the Weyl invariance. The authors of Ref.~\cite{Chamseddine} have shown that this longitudinal mode can
reproduce dark matter and can be considered as a source of the cold dark
matter. Then in ~\cite{Chamseddine1} Chamseddine \emph{et al.} have
investigated the mimetic matter in the presence of an arbitrary
potential $V(\phi)$, by using the Lagrange multipliers approach proposed
in~\cite{Golovnev}. By adopting the appropriate potentials, they
have obtained various cosmological solutions and found that
depending on the choice of the potential, the mimetic matter can
behaves as quintessence, phantom or inflaton fields. Mimetic gravity
in $f(R)$ scenarios leads to interesting results and some authors have studied
its aspects in details. The authors of~\cite{Noj14} have proposed modified mimetic
gravity and studied its early and late time acceleration. They have
also studied the generalization of the model by adding the scalar
potential in Lagrange multiplier framework. In~\cite{Mom14}, the
issue of Noether symmetry for mimetic $f(R)$ gravity has been
investigated. The authors have shown that in this model it is
possible to get bouncing and LCDM solutions. The authors
of~\cite{Shi15} have analyzed the energy conditions and stability of
the mimetic $f(R)$ gravity. Cosmological inflation in mimetic $f(R)$
gravity and its comparison with Planck data has been studied
in~\cite{Odi15}. The unimodular gravity which can potentially
address the cosmological constant problem and late time acceleration
of the Universe has been studied in~\cite{Noj16a} in the light of mimetic
gravity. Other studies in mimetic $f(R)$ gravity can be found
in~\cite{Mat15,Odi16a,Ast16,Odi16b,Ast15,Odi16c,Odi16d,Noj16b}.

With these preliminaries, in this paper we consider a RS II
braneworld model in the spirit of mimetic gravity. We assume the bulk metric is made up of
a non-dynamical scalar field $\Phi$ and an auxiliary metric $\tilde{{\cal{G}}}_{AB}$ so that
${\cal{G}}_{AB}= \tilde{{\cal{G}}}^{CD}\,\Phi_{,C}\,\Phi_{,D}
\,\tilde{{\cal{G}}}_{AB}$ where $A, B, ...$ are the bulk spacetime indices.
We show that the footprint of this scalar field can play the role of a mimetic matter on the brane.
In fact we suppose that the scalar degree of freedom which mimics the dark sectors on the
brane has its origin on a bulk scalar field, $\Phi$.
By writing the effective 4D Einstein equations induced on the brane, we find the effective
Fridmann equation on the brane. By regarding the relation between the 5D
potential and the effective 4D potential and by considering the
effective scalar field on the brane to be $\phi$ (the scalar field defined in
equation (\ref{eq1})), we present the equation for potential of the brane mimetic scalar
field. In this regard, we obtain the Friedmann equation, the scale factor and the equation of
state parameter for some specific potentials in this setup. By numerical analysis on the models
parameters space we investigate the late time accelerating phase of the
universe expansion as well as initial inflation in this setup. We show that it is possible
to realize cosmological inflation in this setup. We show also that equation of state parameter of the
induced mimetic field on the brane crosses the phantom divide line in the same way as observations show:
it evolves from quintessence to phantom phase by crossing the phantom divide in a redshift that is compatible with observation such as Planck2015 data.

\section{Preliminaries}
In mimetic gravity~\cite{Chamseddine,Chamseddine1}, the physical metric is given
by Eq. (\ref{eq1}) where the scalar field satisfies a first order Hamilton-Jacobi type differential equation and
therefore it is not a dynamical field in essence. This scalar field satisfies the constraint $\label{eq2} g^{\mu\nu}\partial_{\mu}\phi\partial_{\nu}\phi=1$.
The action of the model can be written as ~\cite{Chamseddine1}
\begin{equation}\label{eq3}
S=\int d^{4}x \sqrt{-g} \bigg[-\frac{1}{2}R(g_{\mu\nu})+\lambda
\Big(g^{\mu\nu}\partial_{\mu}\phi\partial_{\nu}\phi-1\Big)-V(\phi)+{\cal{L}}_{m}\bigg],
\end{equation}
where $V(\phi)$ is an arbitrary potential of the scalar field, $\lambda$ is a Lagrange multiplier
and ${\cal{L}}_{m}$ is the Lagrangian of the matter fields. The field equations of the model are given by ~\cite{Chamseddine1}
\begin{equation}
\label{eq4}
G_{\mu\nu}-2\lambda\partial_{\mu}\phi\partial_{\nu}\phi-g_{\mu\nu}V(\phi)=T_{\mu\nu}\,.
\end{equation}
Taking the trace of these equations gives the Lagrange multiplier as $\lambda=\frac{1}{2}\Big(G-T-4V\Big)$ so that
\begin{equation}
\label{eq6}G_{\mu\nu}=\Big(G-T-4V\Big)\partial_{\mu}\phi\partial_{\nu}\phi+g_{\mu\nu}V(\phi)+T_{\mu\nu}.
\end{equation}
The equation of motion of the scalar field $\phi$ is ~\cite{Chamseddine1}
\begin{equation}
\label{eq7}\nabla^{\nu}\Big[(G-T-4V)\partial_{\nu}\phi\Big]=-V^{'}(\phi),
\end{equation}
where $V^{'}=\frac{dV}{d\phi}$. By setting the energy-momentum of the scalar field to be of the perfect fluid form, we find $\tilde{p}=-V$ and $\tilde{\varepsilon}=G-T-3V$
where $\tilde{p}$ and $\tilde{\varepsilon}$ are pressure and energy density of the fluid respectively.
For a spatially flat FRW universe with $ds^{2}=dt^{2}-a^{2}(t)\delta_{ik}dx^{i}dx^{k}$ and adopting the hypersurfaces of constant time to be the same as
the hypersurfaces of constant scalar field, from equation
(\ref{eq1}) we obtain $\phi=t$~\cite{Chamseddine1}. Finally, in the absence of ordinary matter the following Friedmann equation gives the cosmological dynamics in this setup
\begin{equation}
\label{eq11}H^{2}=\frac{1}{3}\tilde{\varepsilon}=\frac{1}{a^{3}}\int
a^{2}Vda,
\end{equation}
where $H\equiv\frac{\dot{a}}{a}$ is the Hubble parameter. The amount
of the mimetic dark matter is determined by the constant of
integration in (\ref{eq11}) since it behaves like matter as $\frac{1}{a^{3}}$.
Another mimetic component contributes in (\ref{eq11}) which depends on the types of
various potentials that are used in this setup.

In the next step we focus briefly on the cosmological dynamics of a bulk scalar
field in the RS II braneworld scenario (see Refs. \cite{Him01,Him02,Shi00} for details) to see the relation between bulk and brane potentials.
We assume a positive tension $Z_{2}$ symmetric brane which is embedded in a 5D bulk
with a negative cosmological constant, $\Lambda_{5}$. We set the 5D line element to be as
\begin{equation}\label{eq12}
^{(5)}ds^{2}={\cal{G}}_{AB}\,dx^{A}\,dx^{B}=dy^{2}+g_{\mu\nu}(x^{\alpha},y)dx^{\mu}dx^{\nu},
\end{equation}
where $g_{\mu\nu}$ is the 4D metric induced on the brane ($\mu,\,\nu,\,\alpha =0,1,2,3$) and we assume the brane is
located at $y=0$. The 5D Einstein's equations are given by~\cite{Him01,Shi00}
\begin{equation}
\label{eq13}R_{AB}-\frac{1}{2} {\cal{G}}_{AB}R+\Lambda_{5}{\cal{G}}_{AB}=\kappa^{2}_{5}\bigg(T_{AB}+S_{AB}\delta(y)\bigg),
\end{equation}
where $S_{AB}$ is the energy-momentum tensor on the brane, whereas
$T_{AB}$ stands for the energy-momentum tensor spreading over the
bulk. By assuming a minimally coupled scalar field $\Phi$ in the bulk with
potential $V(\Phi)$, the effective induced 4D Einstein's equations on
the brane are given by~\cite{Him01,Him02,Shi00}
\begin{equation}
\label{eq17}G_{\mu\nu}=\kappa_{4}^{2}T^{(s)}_{\mu\nu}-E_{\mu\nu},
\end{equation}
with $\kappa_{4}^{2}=\frac{\kappa_{5}^{4}\sigma}{6}$,
\begin{equation}
\label{eq19}T_{\mu\nu}^{(s)}=\frac{1}{\kappa_{5}^{2}\sigma}\Bigg[4\Phi_{,\mu}\Phi_{,\nu}
+\bigg(\frac{3}{2}(\Phi_{,y})^{2}-\frac{5}{2}g^{\alpha\beta}\Phi_{,\alpha}\Phi_{,\beta}-3V(\Phi)\bigg)g_{\mu\nu}\Bigg],
\end{equation}
and
\begin{equation}
\label{eq20}E_{\mu\nu}= \,^{(5)}C_{yByD}\,g_{\mu}^{B}\,g_{\nu}^{D}\,,
\end{equation}
where $\sigma$ is the brane tension. $^{(5)}C_{yByD}$ is the 5D Weyl tensor. We consider a spatially isotropic and
homogenous induced metric on the brane as
\begin{equation}
\label{eq21}ds^{2}|_{y=0}=g_{\mu\nu}(y=0)dx^{\mu}dx^{\nu}=-dt^{2}+a^{2}(t)\gamma_{ij}dx^{i}dx^{j},\quad\quad i,j=1,2,3
\end{equation}
where $\gamma_{ij}$ is a maximally symmetric 3D metric with
curvature $k=\pm1,0$. The effective 4D Friedmann
equation on the brane is obtained as~\cite{Him01,Him02}
\begin{equation}
\label{eq22}3\Bigg[\bigg(\frac{\dot{a}}{a}\bigg)^{2}+\frac{k}{a^{2}}\Bigg]\equiv3H^{2}=\kappa_{4}^{2}\rho_{eff},
\end{equation}
where
\begin{equation}
\label{eq23}\rho_{eff}=\frac{3}{\kappa_{5}^{2}\sigma}\bigg(\frac{\dot{\Phi}^{2}}{2}+V(\Phi)\bigg)-\frac{E_{tt}}{\kappa_{4}^{2}}.
\end{equation}
While $E_{tt}$ cannot be determined just by the 4D equations, Bianchi identities are capable to give
some general features of this quantity. In this regard, $E_{tt}$ on the
brane is obtained as~\cite{Him01}
\begin{equation}
\label{eq24}E_{tt}=\frac{\kappa_{5}^{2}}{2a^{4}}\int^{t}_{0}a^{4}\dot{\Phi}\Big(\partial_{y}^{2}\Phi+\frac{\dot{a}}{a}\dot{\Phi}\Big)dt.
\end{equation}
The late time behavior of the bulk scalar field can be derived by analyzing
the asymptotic behavior of the Green function~\cite{Him02}. As an important result for our purpose, the bulk scalar field evaluated on the brane
behaves as an effective 4D scalar field almost identical to the
corresponding system in the standard 4D theory. In this regard,
$E_{tt}$ takes the following form~\cite{Him02}
\begin{equation}
\label{eq25}E_{tt}=-\frac{\kappa_{5}^{2}}{2a^{4}}\int^{t}_{0}a^{4}\dot{\Phi}\bigg(\ddot{\Phi}+2\frac{\dot{a}}{a}\dot{\Phi}\bigg)dt
=-\frac{\kappa_{5}^{2}}{4}\dot{\Phi}^{2}.
\end{equation}
Therefore, the effective energy density is given by
\begin{equation}
\label{eq26}\rho_{eff}=\frac{3}{\kappa_{5}^{2}\sigma}\bigg(\frac{\dot{\Phi}^{2}}{2}+V(\Phi)\bigg)
-\frac{E_{tt}}{\kappa_{4}^{2}}=\frac{1}{2}\dot{\phi}^{2}+V_{eff}(\phi),
\end{equation}
where
\begin{equation}
\label{eq27}\phi=\sqrt{l_{0}}\Phi \quad:\quad
l_{0}=\frac{3}{\kappa_{5}^{2}\sigma},
\end{equation}
and
\begin{equation}
\label{eq28}V_{eff}(\phi)=\frac{l_{0}}{2}V\Big(\frac{\phi}{\sqrt{l_{0}}}\Big).
\end{equation}
After these preliminaries, in the next section we consider the effective scalar
field, which is induced on the brane in this manner, to be the one defined in (\ref{eq1}) as the mimetic field and then we
investigate its both late time and early time cosmological dynamics.

\section{Braneworld Mimetic Cosmology}

\begin{figure}[htp]
\begin{center}\includegraphics{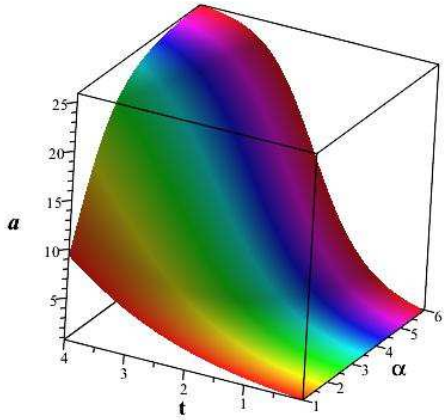} \vspace{4.5cm}
\end{center}
\caption{\label{fig0}\small {Evolution of the scale factor versus $\alpha$ and $t$ with a fixed $l_{0}=0.01$ for $V=V_{0}e^{-\sqrt{\alpha}t}$. }}
\end{figure}

Now we construct a braneworld extension of the mimetic gravity in the
RS II braneworld setup and then we study its cosmological implications. For this purpose,
we isolate the conformal degree of freedom of 5-dimensional gravity in a covariant manner.
As we have mentioned previously, we assume the bulk metric to be made up of
a scalar field $\Phi$ (the conformal degree of freedom of 5-dimensional gravity) and an auxiliary metric $\tilde{{\cal{G}}}_{AB}$ so that
$${\cal{G}}_{AB}= \tilde{{\cal{G}}}^{CD}\,\Phi_{,C}\,\Phi_{,D}
\,\tilde{{\cal{G}}}_{AB}$$ where $A, B, ...$ are the bulk spacetime indices.
We write the action of this mimetic braneworld gravity as follows
\begin{equation}
\label{eq311}S=\int d^{5}x\sqrt{-{\cal{G}}}\Big[-\frac{1}{2\kappa^{2}_{5}}R({\cal{G}}_{AB})+\gamma({\cal{G}}^{AB}\partial_{A}\Phi\partial_{B}\Phi-1)-V(\Phi)+{\cal{L}}_{M}({\cal{G}}_{AB},\ldots)\Big]
\end{equation}
where $\gamma$ is a Lagrange multiplier, $V(\Phi)$ is potential of the scalar field $\Phi$ and  ${\cal{L}}_{M}$ is Lagrangian of other possible fields in the bulk. In this setup, the bulk scalar field (the conformal degree of freedom of 5-dimensional gravity) satisfies the constraint ${\cal{G}}^{AB}\partial_{A}\Phi\partial_{B}\Phi=1$. Einstein field equations of the model are as follows
\begin{equation}
\label{eq312}G_{AB}-2\gamma\partial_{A}\Phi\partial_{B}\Phi-{\cal{G}}_{AB}\,V(\Phi)=\kappa^{2}_{5}T_{AB}\,.
\end{equation}
Taking the trace of these equations gives the Lagrange multiplier as $\gamma=\frac{1}{2}\Big(G-\kappa_{5}^{2}T-5V\Big)$. Therefore, Eq. (\ref{eq312}) can be rewritten as
\begin{equation}
\label{eq313}G_{AB}=(G-\kappa^{2}_{5}T-5V)\partial_{A}\Phi\partial_{B}\Phi+{\cal{G}}_{AB}V(\Phi)+\kappa_{5}^{2}T_{AB}\,.
\end{equation}
Then the equation of motion of the scalar field is as follows
\begin{equation}
\label{eq314}\nabla^{A}[(G-\kappa^{2}_{5}T-5V)\partial_{A}\Phi]=-V'(\Phi)\,.
\end{equation}
Now we set
\begin{equation}
\label{eq315}ds^{2}=-n^{2}(\tau,y)d\tau^{2}+a^{2}(\tau,y)\gamma_{ij}dx^{i}dx^{j}+b^{2}(\tau,y)dy^{2}\,,
\end{equation}
where $\tau$ is a time parameter. If we set ${T}^{A}\,_{B}={T}^{A}\,_{B}|_{bulk}+T^{A}\,_{B}|_{brane}$ with ${T}^{A}\,_{B}|_{bulk}=diag(-\rho_{B},P_{B},P_{B},P_{B},P_{5})$ and
$T^{A}\,_{B}|_{brane}=\frac{\delta(y)}{b}diag(-\rho_{b},p_{b},p_{b},p_{b},0)$, by some appropriate assumptions (which we ignore to state explicitly) the following brane Friedmann equation can be derived
\begin{equation}
\label{eq316}\frac{\dot{a}^{2}_{0}}{a^{2}_{0}}=\frac{\kappa_{4}^{2}}{6}\rho_{B}+\frac{\kappa_{4}^{4}}{36}\rho^{2}_{b}+\frac{C}{a^{4}_{0}}-\frac{k}{a^{2}_{0}}\,,
\end{equation}
where $a_{0}$ is the brane scale factor. If we suppose the only source of the energy-momentum in the bulk to be the scalar field $\Phi$, then $\rho_{B}$ and $P_{B}$ can be obtained easily.
Equation (\ref{eq313}) now takes the following form
\begin{equation}
\label{eq317}G_{AB}=(G-5V)\partial_{A}\Phi\partial_{B}\Phi+{\cal{G}}_{AB}V(\Phi).
\end{equation}
If we set the energy-momentum of the scalar field $\Phi$ to be of the perfect fluid form, we find
\begin{equation}
\label{eq318}P_{B}=-V \qquad\qquad,\qquad\qquad \rho_{B}=G-4V\,.
\end{equation}
Therefore, by setting $\Phi=\tau$, from (\ref{eq314}) we find
\begin{equation}
\label{eq319}\nabla^{0}[(G-5V)\partial_{0}\Phi]=-\frac{dV}{d\Phi}
\end{equation}
which gives
\begin{equation}
\label{eq320}\frac{1}{nba^{3}}\frac{d}{d\tau}\Big(nba^{3}(\rho_{B}-V)\Big)=-\frac{dV}{d\tau}\,.
\end{equation}
By integration we find
\begin{equation}
\label{eq321} nba^{3}(\rho_{B}-V)-C=-\int nba^{3}dV\,,
\end{equation}
where $C$ is a constant we set to be zero. Therefore, we find
\begin{equation}
\label{eq322}\rho_{B}=V-\frac{1}{nba^{3}}\int nba^{3}dV\,.
\end{equation}
An integration by part gives
\begin{equation}
\label{eq323}\rho_{B}=\frac{1}{nba^{3}}\int Vd(nba^{3}).
\end{equation}
If we assume the fifth dimension to be static (that is, $\dot{b}=0$) which enables us to set $b=1$ on the brane and also by a redefinition of the time parameter so that $n|_{brane}=1$,
we find the following familiar (from standard mimetic scenario) relation
\begin{equation}
\label{eq324}H^{2}=\frac{1}{a^{3}}\int Va^{2}da\,.
\end{equation}
We assume that footprint of the bulk scalar field $\Phi$ as induced conformal degree of freedom on the brane plays the role of a mimetic matter on the brane.
In this regard, the mimetic field is induced from the bulk scalar field, $\Phi$. That is, the
mimetic scalar field is the effective 4D scalar field evaluated on
the brane. We assume that the scalar field $\phi$ to be identical
with time $t$ on the brane and there is no contribution from the ordinary matter
fields in the energy-momentum tensor on the brane. By neglecting the non-trivial contribution of bulk Weyl tensor as $E_{tt}$ in
equation (\ref{eq26}) and adopting $\phi=\sqrt{l_{0}}\Phi$, we
obtain the effective energy density of the mimetic field on the brane as follows
\begin{equation}
\label{eq29}\rho_{eff}=l_{0}\Big(\frac{1}{2}+V(\Phi)\Big).
\end{equation}
Now from this equation and also equations (\ref{eq22}) and (\ref{eq324}) we find the potential of the bulk scalar field in terms of
the mimetic potential on the brane as
\begin{equation}
\label{eq30}V(\Phi)=-\frac{1}{2}+\frac{3}{l_{0}\kappa_{4}^{2}a^{3}}\int
a^{2} V(\phi) da\,.
\end{equation}
On the other hand, from equation (\ref{eq28}) we have
\begin{equation}
\label{eq31}V(\Phi)=\frac{2}{l_{0}}V_{eff}\Big(\sqrt{l_{0}}\Phi\Big)\equiv\frac{2}{l_{0}}V\Big(\sqrt{l_{0}}\Phi\Big)=\frac{2}{l_{0}}V(t)\,,
\end{equation}
where $t$ is the brane time coordinate. So, equation (\ref{eq30}) takes the following form
\begin{equation}
\label{eq32}\frac{2}{l_{0}}V(t)=-\frac{1}{2}+\frac{3}{l_{0}\kappa_{4}^{2}a^{3}}\int
a^{2}V(t)da\,.
\end{equation}
By multiplying equation (\ref{eq32}) by $a^{3}$ and differentiating
it with respect to the cosmic time we get
\begin{equation}
\label{eq33} 3a^{2}\dot{a}V+a^{3}\dot{V}=-\frac{3l_{0}}{4}a^{2}\dot{a}+\frac{3}{2\kappa_{4}^{2}}a^{2}
V \dot{a}\,.
\end{equation}
Finally, by using $H\equiv\frac{\dot{a}}{a}$, we obtain
\begin{equation}
\label{eq34}H=\frac{-\dot{V}}{3(1-\frac{1}{2\kappa_{4}^{2}})V+\frac{3l_{0}}{4}}\,.
\end{equation}
From now on, for simplicity we set $\kappa_{4}^{2}=1$ and we investigate the cosmological solutions for some special choices
of the mimetic potential. For the first case, we set
\begin{equation}
V(t)=V_{0}e^{-\sqrt{\alpha}t}\,,
\end{equation}
where $\alpha$ and $V_{0}$ are constants and we set $V_{0}=1$ for simplicity. By solving Eq. (\ref{eq34}), we find
\begin{equation}
\label{scale0}\frac{a}{a_{0}}=\left(\frac{3}{2}\,{{\rm e}^{-\sqrt {\alpha}t}}+\frac{3}{4}\,l_{0} \right) ^{-\frac{2}{3}}\,.
\end{equation}
For small $l_{0}$, corresponding to large brane tension (since $l_{0}=\frac{3}{\kappa_{5}^{2}\sigma}$),
this scale factor turns to $a(t)\sim {\rm e}^{\frac{2}{3}\sqrt {\alpha}t}$
which shows possibility of realization of cosmic inflation for positive $\alpha$ in this setup.
So, this brane mimetic scenario essentially has the capability to realize initial time cosmic inflation. Figure \ref{fig0} shows the behavior of the scale factor (\ref{scale0}) versus cosmic time and the parameter $\alpha$. For large values of $\alpha$, possibility of realization of exponential expansion is evident by the slope of the curves. 

\begin{figure}[htp]
\begin{center}\includegraphics{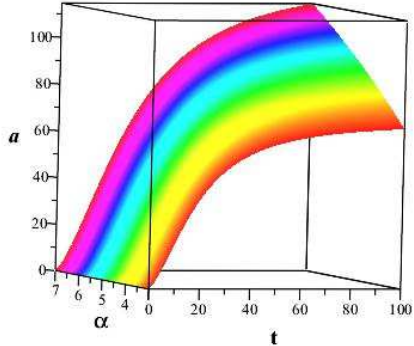} \vspace{4.5cm}
\end{center}
\caption{\label{fig1}\small {Evolution of the scale factor versus $\alpha$ and $t$ with a fixed $l_{0}=0.01$ for $V=\frac{\alpha}{\phi^{2}}$. }}
\end{figure}

As the second case, we adopt the following mimetic potential after Ref.~\cite{Chamseddine1}
\begin{equation}
\label{eq35}V=\frac{\alpha}{t^{2}},
\end{equation}
where $\alpha$ is a constant. Substituting this potential into
equation (\ref{eq34}) we get
\begin{equation}
\label{eq36}\frac{da}{a}=\frac{4}{3}\frac{dt}{t+\frac{l_{0}}{2\alpha}t^{3}}\,,
\end{equation}
By integrating equation (\ref{eq36}), we obtain the scale factor in
this model as follows
\begin{equation}
\label{eq38}\frac{a}{a_{0}}=\Bigg(\frac{t^{2}}{1+\frac{l_{0}}{2\alpha}t^{2}} \Bigg)^{\frac{2}{3}}
\end{equation}
where $a_{0}$ is an integration constant which we re-scale it to
unity. Figure \ref{fig1} shows the behavior of the scale factor versus $\alpha$ and $t$ for a fixed brane tension. For 
sufficiently small time coordinate, corresponding to early universe, $\ddot{a}>0$ which gives a positively accelerated expansion.
So, this model has the potential to realize cosmic inflation at least in some subsets of its parameter space. A simple calculation
shows also that for sufficiently large $\alpha$ and small $l_{0}$ (that is, large brane tension), the scale factor tends to $a(t)\sim t^{\frac{4}{3}}$ which gives an accelerating expansion.

\begin{figure}[htp]
\begin{center}\includegraphics{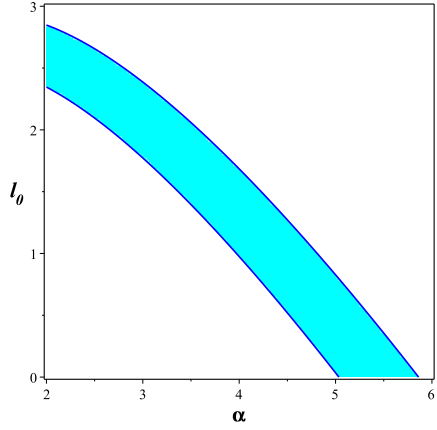} \vspace{4.5cm}
\end{center}
\caption{\label{fig2}\small {The ranges of the parameters $\alpha$ and
$l_{0}$ for $\omega=-1.019^{+0.075}_{-0.080}$ at
present time for $V=\frac{\alpha}{\phi^{2}}$.}}
\end{figure}

The equation of state parameter with potential as Eq.~(\ref{eq35}) in this mimetic braneworld setup is given by
\begin{equation}
\label{eq39}\omega=-\Big(\frac{3\alpha}{16}\Big)\Big[1+\frac{l_{0}}{2\alpha}t^{2}\Big]^{2}\,.
\end{equation}
To seek for late time acceleration in this setup, we perform some
numerical analysis on the model's parameters space. Based on the Planck2015 observational
data~\cite{planck2015de} the current value of the equation of state
parameter is constraint as $\omega=-1.019^{+0.075}_{-0.080}$. With this point in mind,
by numerical study of $\omega$ (defined by equation (\ref{eq39})) we
obtain the ranges of the parameters $\alpha$ and $l_{0}$ compatible with the constraint on the equation of state parameter from Planck2015 data set. The
result is shown in figure \ref{fig2}. Note that $l_{0}$ is related to the brane tension via $l_{0}=\frac{3}{\kappa_{5}^{2}\sigma}$.

\begin{figure}[htp]
\begin{center}\includegraphics{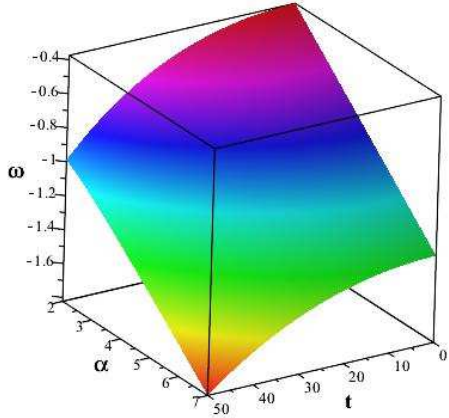}
 \vspace{5cm}
\end{center}
\caption{\label{fig3}\small {Evolution of the equation of state
parameter versus the cosmic time and $\alpha$ with a fixed $l_{0}=0.01$ for $V=\frac{\alpha}{\phi^{2}}$.}}
\end{figure}

To have an accelerating expansion, the effective equation of state
parameter should be less than $-\frac{1}{3}$. On the other hand, the
equation of state parameter is a dynamical parameter which its
value changes by evolution of the universe. According to the
observational data, $\omega$ has crossed the phantom divide line ($\omega=-1$) at the
near past. In fact, observations show that the universe had a transition
from a quintessence phase to a phantom phase. So, a successful dark energy model
should realize a crossing of the phantom divide in the past. In this regard, we study the evolution of
$\omega$ versus the the cosmic time to see its capability to realize a phantom divide crossing. The
results are shown in figures \ref{fig3}. As this figure shows, depending on the value of $\alpha$ 
(specially for sufficiently small $\alpha$ such as $\alpha=3$), the equation of state parameter in this model starts
from $\omega>-1$ and then crosses the phantom divide at a redshift which depends on the value of $\alpha$.
So, this model is successful to address the late time cosmic dynamics in a fascinating manner.

\begin{figure}[htp]
\begin{center}\includegraphics{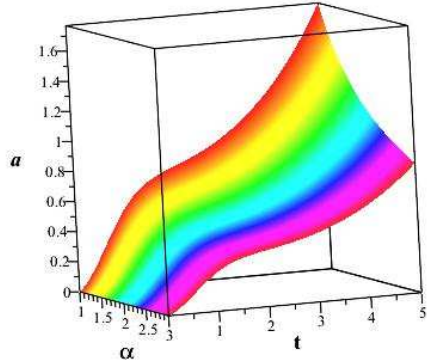} \vspace{5cm}
\end{center}
\caption{\label{fig4}\small {Evolution of the scale factor versus $\alpha$ and $t$ with a fixed $l_{0}=0.01$ for $V(\phi)=\frac{\alpha\phi^{2}}{1+e^{\sqrt{\alpha}\phi}}+\frac{\beta}{\phi^{2}}$. }}
\end{figure}

As the third and even richer example, we consider the following potential
\begin{equation}
V=\frac{\alpha\phi^{2}}{1+e^{\sqrt{\alpha}\phi}}+\frac{\beta}{\phi^{2}}\,,
\end{equation}
where with $\phi=t$ gives the following scale factor
\begin{equation}
\frac{a}{a_{0}}={\frac {{t}^{4/3} \left( 1+{{\rm e}^{\sqrt
{\alpha}t}} \right) ^{2/3}} { \left(
2\,\alpha\,{t}^{4}+l_{0}\,{t}^{2}{{\rm e}^{\sqrt {\alpha}t}}+l_{0}\,{t}^{
2}+2\,\beta\,{{\rm e}^{\sqrt {\alpha}t}}+2\,\beta \right) ^{2/3}}}\,.
\end{equation}
For simplicity in our numerical analysis, in what follows we set $\alpha=\beta$. Figure \ref{fig4} shows the behavior of this scale factor.
As this figure shows, the adopted potential has the capability to realize initial inflation
as well as the late time accelerated expansion. These features can be seen via the slope of the curves which depends on the value of $\alpha=\beta$.
The equation of state parameter in this case is as follows

\begin{equation}
\omega=-\frac{1}{3}\, \left( {\frac {\alpha\,{t}^{2}}{1+{{\rm
e}^{t}}}}+{\frac {\beta }{{t}^{2}}} \right)  \left(
\frac{3}{2}\,{\frac {\alpha\,{t}^{2}}{1+{{\rm e}^{ \sqrt
{\alpha}t}}}}+\frac{3}{2}\,{\frac {\beta}{{t}^{2}}}+\frac{3}{4}\,l_{0}
\right) ^{2} \left( {\frac {2\alpha\,t}{1+{{\rm e}^{\sqrt
{\alpha}t}}}}-{\frac { {\alpha}^{3/2}{t}^{2}{{\rm e}^{\sqrt
{\alpha}t}}}{ \left( 1+{{\rm e}^{ \sqrt {\alpha}t}} \right)
^{2}}}-{\frac {2\beta}{{t}^{3}}} \right) ^ {-2}\,.
\end{equation}

\begin{figure}[htp]
\begin{center}\includegraphics{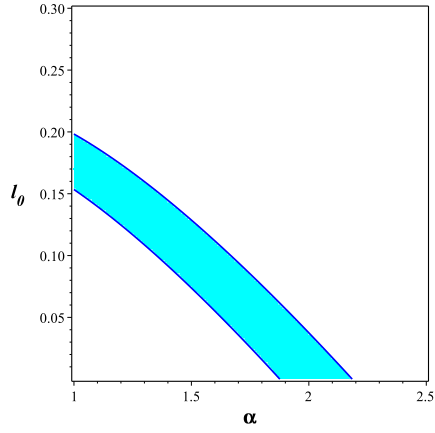} \vspace{3cm}
\end{center}
\caption{\label{fig5}\small {The ranges of the parameters $\alpha$ and
$\beta$ for $\omega=-1.019^{+0.075}_{-0.080}$ at
present time for $V(\phi)=\frac{\alpha\phi^{2}}{1+e^{\sqrt{\alpha}\phi}}+\frac{\beta}{\phi^{2}}$.}}
\end{figure}

Figure \ref{fig5} gives the constraint on $\alpha$ and $l_{0}$ from the Planck2015 observational
data in this case.

\begin{figure}[htp]
\begin{center}\includegraphics{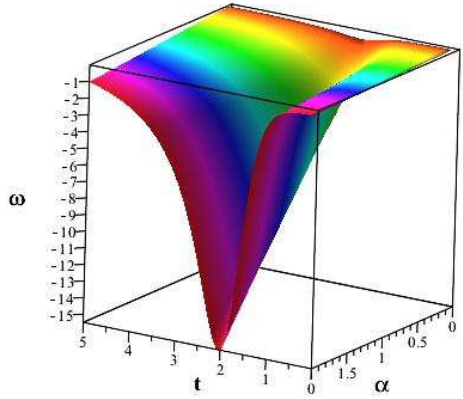}
 \vspace{3cm}
\end{center}
\caption{\label{fig6}\small {Evolution of the equation of state
parameter versus the cosmic time and $\alpha$ with a fixed $l_{0}=0.01$ for $V(\phi)=\frac{\alpha\phi^{2}}{1+e^{\sqrt{\alpha}\phi}}+\frac{\beta}{\phi^{2}}$.}}
\end{figure}

Finally, figure \ref{fig6} gives the evolution of the equation of state parameter. As this figure shows,
late time cosmic acceleration and phantom divide crossing can be addressed in this case successfully. 
It is interesting to note that the case with $\omega=-1$ as a late time cosmological dominated universe is well in the parameter space of the model.

\section{Summary and Conclusion}

This work has been devoted to an extension of the idea of mimetic gravity to braneworld scenario.
In the original mimetic matter scenario, Chamseddine and Mukhanov have formulated 4D Einstein's theory of gravity by
isolating the conformal degree of freedom in a covariant manner~\cite{Chamseddine}. They have introduced a physical metric
defined in terms of an auxiliary metric and a scalar field appearing through its first derivatives.
Then they have shown that the conformal degree of freedom becomes dynamical even in the absence of matter
and this mimetic field has the potential to be a candidate for dark matter. They have proposed minimal extensions of mimetic matter scenario
by introducing a potential for mimetic scalar field to explain several important issues such as cosmological inflation, quintessence and bouncing nonsingular universe.
In this paper we have extended the idea of mimetic gravity to a barneworld scenario. 
For this purpose, we have isolated the conformal degree of freedom for 5D gravity in a covariant manner. 
We have assumed that the bulk metric is made up of a scalar field $\Phi$ and an auxiliary metric $\tilde{{\cal{G}}}_{AB}$ so that
${\cal{G}}_{AB}= \tilde{{\cal{G}}}^{CD}\,\Phi_{,C}\,\Phi_{,D}
\,\tilde{{\cal{G}}}_{AB}$. Then we have shown that the induced conformal degree of freedom on the brane as induced scalar field can play the role of a mimetic matter on the brane. In fact we have supposed that the scalar degree of freedom which mimics the dark sectors on the
brane has its origin on a bulk scalar field, $\Phi$. By projecting the bulk field equations on the brane we have studied cosmological implications of this extended mimetic scenario. By adopting some potentials we have shown that this brane mimetic scenario explains initial cosmic inflation as well as the late time positively accelerated expansion. Specially, we have shown that by adopting a potential of the type $V(\phi)=V_{0}e^{-\sqrt{\alpha}\phi}$, for small $l_{0}$, corresponding to large brane tension,
the scale factor becomes as $a(t)\sim {\rm e}^{\frac{2}{3}\sqrt {\alpha}t}$
which shows possibility of realization of cosmic inflation for positive $\alpha$ in this setup. 
We have shown also that this mimetic braneworld scenario explains late time cosmic dynamics in a fascinating manner: the universe has entered in a positively accelerated phase of expansion in near past with an equation of state parameter for mimetic field that depending on the value of parameter $\alpha$ crosses the phantom divide line ($\omega=-1$) from quintessence to phantom phase with a redshift well in the range of observational data. By adopting the mimetic potential as
$V(\phi)=\frac{\alpha}{\phi^{2}}$ and also $V(\phi)=\frac{\alpha\phi^{2}}{1+e^{\sqrt{\alpha}\phi}}+\frac{\beta}{\phi^{2}}$\,, we have constraint the model parameters by confrontation with Planck2015 data.\\

{\bf Acknowledgement}\\

We thank Dr Narges Rashidi for insightful comments and careful reading of the manuscript.

\end{document}